\documentclass[11pt]{article}
\usepackage[pdfstartview=FitH,colorlinks=true,linkcolor=blue,anchorcolor=red,citecolor=magenta,urlcolor=blue]{hyperref}
\usepackage[english]{babel}

\usepackage{amsmath,amssymb,titling,authblk}
\usepackage{amscd}

\usepackage{amsfonts}
\usepackage{amstext}
\usepackage{amsthm}

\usepackage[normalem]{ulem}
\usepackage{appendix}
\usepackage{bbold}
\usepackage{yfonts}
\usepackage{epsfig}
 \usepackage{microtype}
 
 \usepackage[utf8]{inputenc}
\usepackage[T1]{fontenc}
\usepackage {color}
\usepackage{bbold}

\usepackage{latexsym}
\usepackage{soul}
\setstcolor{blue}
\usepackage{comment}
\usepackage{xcolor}
\definecolor{myurlcolor}{rgb}{0,0,0.4}
\definecolor{mycitecolor}{rgb}{0,0.5,0}
\definecolor{myrefcolor}{rgb}{0.5,0,0}

\usepackage{cite}

\allowdisplaybreaks[4]
\numberwithin{equation}{section}

\definecolor{verde}{cmyk}{.83,.21,1,.08}

\newtheorem{theorem}{Theorem}

\newtheorem*{proof*}{Proof}

\advance\hoffset by -1.1truecm
\advance\voffset by -1.0truecm
\newcommand{\be}{\begin{equation}}
\newcommand{\ee}{\end{equation}}
\newcommand{\beqa}{\begin{eqnarray}}
\newcommand{\eeqa}{\end{eqnarray}}
\newcommand{\eqn}[1]{(\ref{#1})}

\newcommand{\del}{\partial}

\newcommand{\R}{\mathrm R}

\numberwithin{equation}{section}

\newcounter{appendice}

\usepackage{lmodern}

\begin{document}

%\setlength{\droptitle}{-6pc}
%\pretitle{\begin{flushright}\small
%ICCUB-\ldots
%\end{flushright}\vspace*{2pc}%
%\begin{center}\LARGE}
%\posttitle{\par\end{center}}

\title{Symplectic realizations and Lie groupoids in Poisson Electrodynamics}

\renewcommand\Affilfont{\itshape}
\setlength{\affilsep}{1.5em}

\author[1,2]{Fabio Di Cosmo\thanks{fcosmo@math.uc3m.es }}
\author[1,2]{Alberto Ibort\thanks{albertoi@math.uc3m.es}}
\author[3,4]{Giuseppe Marmo\thanks{giuseppe.marmo@na.infn.it}}
\author[3,4]{Patrizia Vitale\thanks{patrizia.vitale@na.infn.it}}
\affil[1]{Department of Mathematics, University Carlos III de Madrid, Leganés, Madrid, Spain}
\affil[2]{ICMAT, Instituto de Ciencias Matemáticas (CSIC-UAM-UC3M-UCM)}
\affil[3]{Dipartimento di Fisica ``Ettore Pancini'', Universit\`{a} di Napoli {\sl Federico~II}, Napoli, Italy}
\affil[4]{INFN, Sezione di Napoli, Italy}

\date{}

\maketitle

\vspace{-2cm}

\begin{abstract}
We define the gauge potentials of Poisson electrodynamics as sections of a symplectic realization of the spacetime manifold and infinitesimal gauge transformations as a representation of the associated  Lie algebroid acting on the symplectic realization. Finite gauge transformations are obtained by integrating the sections  of the Lie algebroid to bisections of a  symplectic groupoid, which form a one-parameter group of transformations, whose action on the fields of the theory is realized in terms of an action groupoid. A covariant electromagnetic two-form is  obtained, together with a dual two-form, invariant under gauge transformations. The duality appearing in the picture originates from the existence of a pair of orthogonal foliations of the symplectic realization, which produce dual quotient manifolds, one related with space-time, the other with momenta. 
 \end{abstract}

\section{Introduction}
In this paper we elaborate on a recent approach to Poisson electrodynamics \cite{Kupriyanov:2020sgx, Kupriyanov:2020axe, Kupriyanov:2021aet, Kurkov:2021kxa, Kupriyanov:2021cws,  Kupriyanov:2022ohu, Abla:2022wfz, Kupriyanov:2023gjj}, by proposing an interpretation of gauge fields and gauge transformations in terms of symplectic realizations and groupoids. A similar proposal  has  already been formulated  in \cite{Kupriyanov:2023qot}; we shall comment below on the similarities and differences between the two approaches.  

To start with,   given a non-trivial Poisson structure, $\Theta$,  on the spacetime manifold $M$,   gauge potentials, $A$,  are identified with  the sections of a symplectic realization $\mathcal{U}\subset (T^*M, \omega)$, with $\Lambda=\omega^{-1}$ a Poisson tensor on $\mathcal{U}$ which projects  to $\Theta$. The electromagnetic field is then obtained as a two-form on $M$ as the pull-back of the symplectic form $\omega$ through the potential one-form, $F=A^*(\omega)$. 

Infinitesimal gauge transformations of the gauge potential are  expressed as 
vertical vector fields along the image of the section $A$; we shall see that they are generalised Lie derivatives associated with bundle isomorphisms.  Then, infinitesimal gauge transformations of the electromagnetic field are given in terms of vector fields which are tangent to the vector bundle of two-forms on $M$; these are proper Lie derivatives along the Hamiltonian vector fields $Y_f=\Theta(\mathrm{d}f)$. 

In the second part of the paper we introduce Lie algebroids and Lie groupoids. A similar approach  has  been already proposed  in \cite{Kupriyanov:2023qot}, directly at the level of finite gauge transformations, which are there identified with bisections of the symplectic groupoid integrating the starting Poisson manifold, whereas the gauge potentials are identified with the Lagrangian bisections of the symplectic groupoid. In our case, as we shall see, the relevant structure at the level of spacetime  is the symplectic realization, whereas the groupoidal structure only  matters  at the level of gauge transformations.  We start from the infinitesimal setting: we  recognize the infinitesimal gauge transformations  introduced before as a representation of the  Lie algebroid $(T^*M, a, [\;,\;]_\Theta)$  on the sections of the bundle $\mathcal{U}\rightarrow M$ obtained by the symplectic realization. \footnote{$[\;,\;]_\Theta$ and $a$ are respectively the Lie bracket and the anchor map defined in Eq. \eqn{cotgalgebroid} } The exponential map  of the  algebroid associates with locally closed sections of the form $\mathrm{d}f$  a one-parameter group of bisections of the symplectic groupoid $\Sigma(M)$, say $\mathcal{G}_t=\{\mathrm{Exp}(t \mathrm{d}f)\}$. A one-parameter group of diffeomorphisms of the symplectic realization $\mathcal{U}$  is thus obtained by considering the pull-back of $\mathcal{G}_t$ to  the so-called action groupoid $\Sigma(M)\star\mathcal{U}$, which in turn encodes the action of a groupoid $G\rightrightarrows M$ on the bundle $\mathcal{U}\rightrightarrows M$. 
The infinitesimal generators of this action are precisely the vertical vector fields of gauge transformations found in the first part of the paper. 

As for the electromagnetic field, besides the definition given in terms of  the pull-back of the symplectic form through the section $A$, we shall see that it is possible to define a dual object, which is invariant under gauge transformations, as opposed to the previous one, which is instead covariant. Indeed the symplectic realization $\mathcal{U}$ is endowed with a pair of foliations, the first one is generated by vector fields which are tangent to the leaves of  the bundle map $\pi:\mathcal{U}\rightarrow M$ while the other is defined in terms of vector fields which are symplectic orthogonal to the previous one, with a new bundle map $\pi_1: \mathcal{U}\rightarrow M$. The sections of this second bundle over $M$, say $A^{(1)}$, will define a different two-form $F^{(1)}= {A^{(1)}}^*(\omega)$. The latter can be seen to be invariant under gauge transformations. The covariant and invariant tensor fields, $F$ and $F^{(1)}$ coincide with those obtained in \cite{Kupriyanov:2023qot}  as the   pullback of the symplectic form through the two maps determined by any bisection $\sigma$ of the symplectic groupoid $\Sigma(M)$. 

Our results are limited to local symplectic realizations $\mathcal{U}\subset T^*M$. In order to build a complete theory, a complete symplectic realization is needed. According to \cite{Crainic-fernandes_2} complete symplectic realizations of Poisson manifolds exist if and only if the Poisson manifold is integrable, namely, the Lie algebroid $(T^*M, a, [\;,\;]_\Theta)$ integrates to the associated symplectic groupoid. This has to be checked case by case.

The paper is organised as follows. In Sec. \ref{Poisson Electrodynamics} we review Poisson electrodynamics in terms of symplectic realizations of the space-time manifold and Hamiltonian vector fields associated with gauge transformations. In Sec. \ref{Lie Groupoids} we introduce Lie groupoids and Lie algebroids. We shall need in particular the concept of { bisections} of an  Action Groupoid described in \ref{Action groupoids}. The latter encodes the action of a Lie groupoid $G\rightrightarrows M$ on a bundle with base $M$, whereas the bisections associated with sections of Lie algebroids are the appropriate objects which  describe finite  gauge transformations.  
In Sec. \ref{back to ED} we go back to the model discussed  in Sec. \ref{Poisson Electrodynamics} and interpret the results in terms of the geometric picture which we have developed. In particular, the gauge fields are identified with sections of a bundle over $M$ furnished by the symplectic realization, which carry a representation of the Lie algebroid $T^*M$, that can be exponentiated  to a symplectic groupoid. Finite gauge transformations are obtained as bisections of the associated action groupoid. Finally, in Sec. \ref{conclusions} we draw some conclusions and discuss the results in view of future developments.

\section{Poisson electrodynamics revisited}\label{Poisson Electrodynamics}
In this section we review previous results on Poisson Electrodynamics, recently introduced in \cite{Kupriyanov:2020sgx, Kupriyanov:2020axe, Kupriyanov:2021aet, Kurkov:2021kxa, Kupriyanov:2021cws,  Kupriyanov:2022ohu, Abla:2022wfz, Kupriyanov:2023gjj} adopting a  point of view which will lead us to a geometric reformulation of the model in the second part of the paper. 

In the first part, the main focus will be on the definition of local potentials for the electromagnetic field and their gauge transformations when the space-time $M$ is endowed with a Poisson structure, $\Theta$. The latter is  interpreted as a semi-classical limit of  non-commutative space-time. As in \cite{Kupriyanov:2020sgx, Kupriyanov:2020axe, Kupriyanov:2021aet, Kurkov:2021kxa, Kupriyanov:2021cws,  Kupriyanov:2022ohu, Abla:2022wfz, Kupriyanov:2023gjj}, a  guiding principle  will be the reduction to  classical electrodynamics when the noncommutative shadow, the Poisson structure on the spacetime, vanishes. 

A second feature of the description consists of a local symplectic realization \cite{Weinstein83, Crainic11} of a Poisson manifold $(M,\Theta)$ in terms of an open submanifold $\mathcal{U}$ of the cotangent bundle $T^*M$. The latter is  endowed with an appropriate symplectic form, $\omega$, such that  the projection $\pi \, \colon\, T^*M \,\rightarrow\,M$ is a Poisson map from $(T^*M,\Lambda=\omega^{-1})$ to $(M,\Theta)$. 
Accordingly, the gauge theory  will be defined in terms of potentials for the electromagnetic field, regarded as local sections of the symplectic manifold $(T^*M,\omega)$. Let us remark that the symplectic structure $\omega$ on $T^*M$ is not the canonical symplectic structure on the cotangent bundle of $M$, the latter being denoted by $\omega_0$.

\subsection{The electromagnetic field}
In the standard description of electrodynamics the electromagnetic field is described by a 2-form on the spacetime while the potentials are locally described by one-forms. A natural observation  is  that, if  a local symplectic realization of $(M,  \Theta)$ is available, it is possible to deform the definition of the standard electromagnetic field, of the corresponding potentials and the algebra of gauge transformations in a way that is coherent with the underlying Poisson geometry. 

In order to illustrate the model, let us start from the definition of the electromagnetic field in the standard situation where $M$ has a trivial Poisson structure and $T^*M$ is the canonical cotangent space with symplectic form $\omega_0$. Neglecting possible topological obstructions, the electromagnetic field is described by a closed 2-form $F$ on $M$ and the gauge potentials  are 1-forms on $M$ such that $F = \mathrm{d}A$. Other than being the cotangent bundle of  $M$, $(T^*M, \omega_0)$ gives   a canonical symplectic realization of the Poisson manifold $(M, \Theta=0)$.  Therefore,  the same fields $A$ and $F$ can be recovered as follows: take a section $A\,\colon\,M\,\rightarrow\,T^*M$ of  $(T^*M,\omega_0)$ seen as a symplectic realization of $M$ and define the 2-form $F(A)=A^*(\omega_0)$. Since $\omega_0$ is the differential of the tautological 1-form $\lambda_0$, we have that 
\begin{equation}
    F(A)=A^*(\omega_0) = \mathrm{d} A^*(\lambda_0) = \mathrm{d}A = F\,, 
\end{equation}
which, in Darboux coordinates ${\lbrace x^{\mu},\,p_{\mu} \rbrace}_{\mu=1,\cdots, 4}$, $\omega_0= d p_\mu\wedge dx^\mu$, 
reads 
$$
F_0= A^*(\omega_0)= d A_\mu \wedge dx^\mu= \frac{1}{2}\left(\del_\nu A_\mu-\del_\mu A_\nu\right) dx^\nu\wedge dx^\mu
$$
so to obtain the standard 2-form describing the electromagnetic field. Moreover, since potentials of the symplectic structure are defined up to addition of closed 1-form, which is locally  the differential of a function $f\,\colon\,T^*M\,\rightarrow\,\mathbb{R}$, we derive the fact that gauge potentials are not uniquely determined, and the action of a group of gauge transformations can be included.

When the space-time $M$ is endowed with a non-trivial Poisson structure, we can generalize the previous construction in a natural way. Indeed, if we consider a local symplectic realization $\mathcal{U}\subset (T^*M,\omega)$ of the Poisson manifold $(M,\Theta)$, we can associate again the sections of this bundle with gauge potentials and obtain a deformed electromagnetic field by pulling back the deformed symplectic form $\omega$ on $T^*M$. That is, if $A\,\colon\,U\subset M\,\rightarrow\,T^*M$ is a local section, we can write:
\begin{equation}
    F(A) = A^*(\omega)\,
    \label{eq_1}
\end{equation}
which is  a closed 2-form on the spacetime $M$.

Symplectic realizations of Poisson manifolds are not unique. For example, when the spacetime $(M,\Theta)$ can be identified with  the dual of a Lie algebra, $\mathfrak{g}^*$, with $\Theta$ the Kirillov bracket  $\Theta(x)= c^{ij}_k x^k$ ($c^{ij}_k$ being the structure constants of the algebra $\mathfrak{g}$), a natural symplectic realization is represented by  $T^*G$, the cotangent bundle of the Lie group $G$. But it is well known that  smaller  symplectic realizations of $\mathfrak{g}^*$ are also possible (see for example \cite{Weinstein83, Manko:1993dp, Gracia-Bondia:2001ynb}). 

For the present analysis it is important that the symplectic realization be contained in the cotangent bundle and that the base manifold be spacetime itself. The existence of such a structure  is proven in \cite{Weinstein83, Crainic11}.

Shortly, a symplectic realization on the cotangent bundle $T^*M$ of the Poisson bracket $\Theta$ is constructed in \cite{Crainic11} as follows. One considers the canonical symplectic form $\omega_0$ on $T^*M$ and a Poisson spray, namely, a vector field $V_{\Theta}$ on $T^*M$ such that:
\begin{itemize}
    \item $\pi_*(V_{\Theta}(\xi)) = \Theta(\xi,\cdot)$, $\forall \xi \in T^*M$, where $\pi\,\colon\,T^*M\,\rightarrow M$ is the natural projection of the cotangent bundle;
    \item $m_t^*(V_{\Theta}) = tV_{\Theta}$ where $m_t\,\colon\,T^*M\,\rightarrow\,T^*M$ is the multiplication by non-zero real numbers along the fibres of $T^*M$. This amounts to say that $V_{\Theta}$ is homogeneous of degree 1. 
\end{itemize}
Then one can prove  the following \cite{Crainic11}:
\begin{theorem}
Given a Poisson manifold $(M,\Theta)$ and a Poisson spray $V_{\Theta}$, then there exists an open neighborhood $\mathcal{U}\subset T^*M$ of the zero section so that 
\be
\omega:= \int_0^1 (\phi_t^\Theta)^*\omega_0 \;\mathrm{d}t
\ee
is a symplectic structure  on $\mathcal{U}$ and the canonical projection $\pi\,\colon\,\mathcal{U}\,\rightarrow\,M$ is a symplectic realization.
\end{theorem}
In the previous formula  $\phi_t^{\Theta}\,\colon\,T^*M\,\rightarrow\,T^*M$  denotes the flow of the Poisson spray. It is interesting to notice that there are two dual foliations associated with this symplectic realization: one,  denoted by $\mathcal{F}(\pi)$, is given by the vector fields which are tangent to the leaves of the submersion $\pi$, the other, denoted by   $\mathcal{F}(\pi_1)$,  is given by the vector fields tangent to the leaves of the submersion $\pi_1 = \pi\circ \phi_1^\Theta$. In particular, these two foliations are symplectic orthogonal, namely  $\omega(X, Y) = 0$ for every $X\in \mathcal{F}(\pi)$ and $Y\in \mathcal{F}(\pi_1)$.

In the case of an open manifold $\mathcal{O}\subset \mathbb{R}^n$ one can choose the simplest Poisson spray $V_{\Theta}= \Theta^{\mu\nu}p_\mu \del_\nu$, and the symplectic form $\omega$ can be expressed in Darboux coordinates as
$
\omega= dy^\mu \wedge dp_\mu 
$ where  $y^\mu (x, p)$ is defined in terms of the spray according to  $y^\mu (x, p)= \int_0^1 x^\mu \circ \phi^\Theta_t dt$. The Jacobian matrix $\varepsilon(x,p)= (\del y^\mu/\del x^\nu)$ is formally invertible. 
The associated Poisson bracket, $\Lambda= \omega^{-1}$, reads then
\be\label{Lambd}
\Lambda= \Theta^{\mu\nu}(x)\frac{\del}{\del x^\mu}\wedge \frac{\del}{\del x^\nu}+ \gamma^{\mu}_{\nu}(x,p)\frac{\del}{\del x^\mu}\wedge \frac{\del}{\del p_\nu}
\ee
with $\Theta^{\mu \nu}(x)\partial_{\mu}\wedge \partial_{\nu}$  the local expression of the Poisson bivector $\Theta$ and $\gamma(x,p)$ the inverse matrix of $\varepsilon (x,p)$. $\Lambda$ satisfies the projection condition $\pi_\star\Lambda = \Theta$ by construction.\footnote{
A related approach \cite{Kupriyanov:2021aet} consists in assuming
$\Lambda$ to be of the form \eqn{Lambd} and imposing Jacobi identity. This yields and equation for $\gamma$, 
\beqa
&& \gamma^{ \mu}_{ \nu} \frac{\del}{\del p_\nu } \gamma^\rho_\sigma - \gamma^{ \rho}_{ \nu} \frac{\del}{\del p_\nu } \gamma^\mu_\sigma+ \Theta^{ j  i } \partial_{ i } \gamma^{ k}_{ l} 
- \Theta^{ k i } \partial_{ i } \gamma^{ j }_{ l} 
- \gamma^{ i }_{ l}\partial_{ i }\Theta^{ j  k} = 0,\label {first} \\
&& \gamma^\mu_\nu \,\stackrel{\Theta\rightarrow 0}{\longrightarrow} \,\delta^\mu_\nu\nonumber .
\eeqa}

Eq. \eqn{Lambd}
can be inverted to give an explicit expression for the symplectic form $\omega$, which reads,  
\begin{equation}
    \omega = \varepsilon^{\mu}_{\nu}(x,p) \mathrm{d}p_{\mu}\wedge\mathrm{d}x^{\nu} + \frac{1}{2} \varepsilon^{\mu}_{\alpha}(x,p) \Theta^{\alpha \beta}(x) \varepsilon^{\nu}_{\beta} (x,p)\mathrm{d}p_{\mu}\wedge\mathrm{d}p_{\nu} .    
    \label{eq_2}
\end{equation}

In what follows we will focus on the case of an open manifold with the nondegenerate Poisson tensor given by  Eq.$\eqref{Lambd}$: this will allow for a clear  geometric interpretation of the results obtained in  \cite{Kupriyanov:2021aet}.  It is possible to extend the analysis to the open neighborhood $\mathcal{U}$ of the zero section of $T^*M$, although this  implies that the description is still an infinitesimal one.

Having obtained the symplectic form \eqn{eq_2}, we apply the 
 prescription \eqn{eq_1}. The   electromagnetic field on the Poisson manifold $(M, \Theta)$  reads then
\begin{equation}
    F(A)= \frac{1}{2}\left( \varepsilon^{\mu}_{\sigma}(x,A) \partial_{\rho} A_{\mu} - \varepsilon^{\mu}_{\rho}(x,A) \partial_{\sigma} A_{\mu} + \varepsilon^{\mu}_{\alpha}(x,A) \Theta^{\alpha \beta}(x) \varepsilon^{\nu}_{\beta}(x,A) \partial_{\rho} A_{\mu} \partial_{\sigma} A_{\nu} \right) \mathrm{d}x^{\rho}\wedge\mathrm{d}x^{\sigma}\,,
\end{equation}
thus exhibiting an explicit dependence  on the non-commutativity of the space-time. 
As desired, it  correctly reproduces  the standard electromagnetic field when the Poisson structure on $M$ vanishes (i.e. $\Theta\rightarrow 0$ and $\varepsilon^{\mu}_{\nu}\rightarrow \delta^\mu_\nu$). Let us remark that this expression for $F$  is slightly different from the one obtained {\it e.g.} in  %\cite{Kupriyanov:2020sgx, Kupriyanov:2020axe, Kupriyanov:2021aet, Kurkov:2021kxa, Kupriyanov:2021cws,  Kupriyanov:2022ohu, 
\cite{Kupriyanov:2022ohu}, although retaining the same structure. 

From the local expression one also notices that the electromagnetic field is not a linear functional of the gauge potential, a characteristic  that already occurs in standard non-Abelian gauge theories.   

As a further remark, we notice that, being $F$ defined in terms of the 2-form $\omega$ on the cotangent bundle, there is room for a larger group of gauge transformations. This will be discussed in the next subsection. 

\subsection{Gauge transformations}\label{Gauge transformations 1}
The electromagnetic  field has been obtained as  the  pull-back of the symplectic form $\omega$. It is therefore  natural to  look for   gauge transformations as a  subgroup of the group of symplectomorphisms of $(T^*M,\omega)$. More precisely, in order to recover the standard gauge transformations in the limit of vanishing $\Theta$, one has to restrict to   symplectomorphisms which project onto Poisson morphisms of $(M,\Theta)$. 
Therefore, the  generators of infinitesimal  transformations are the Hamiltonian vector fields 
\begin{equation}\label{Xf}
X_f = \Lambda(\mathrm{d}\pi^*(f),\cdot)\,,
\end{equation}
where $f\in \mathcal{F}(M,\R)$ are the gauge parameters. %$\,\colon\,M\,\rightarrow\,\mathbb{R}$ is a function on $M$, the gauge parameter. 
In local coordinates Eq. (\ref{Xf}) reads
\be\label{localX}
X_f= \Theta^{\mu\nu}\frac{\del f}{\del x^\mu}\frac{\del}{\del x^\nu}+ \gamma^\mu_\nu \frac{\del f}{\del x^\mu}\frac{\del}{\del p_\nu}.
\ee 
This formula shows that, for “Casimir functions” of the space-time Poisson structure $\Theta$,
 $X_f$  are   purely vertical  and project onto the null vector field on M.
The vector fields $X_f$  are $\pi$-related to Hamiltonian vector fields on $M$
\be
\label{hamiltonian vector field Xf}
Y_f = \Theta(\mathrm{d}f,\cdot) ,  
\ee
namely, $\pi_*(X_f)= Y_f$.
In local coordinates we have 
\be\label{localY}
Y_f= \Theta^{\mu\nu}\frac{\del f}{\del x^\mu}\frac{\del}{\del x^\nu}.
\ee
Therefore, when the Poisson structure on $M$ vanishes, we recover the standard vertical translations of the vector bundle $T^*M$. 

Being  $X_f$ Hamiltonian vector fields,   they satisfy the following Lie algebra relations
\begin{equation}
    \left[ X_f\,,\,X_g \right] = X_{\left\lbrace f\,,\,g\right\rbrace_{\Theta}}\,,
\end{equation}
where $\left\lbrace f\,,\,g\right\rbrace_{\Theta}$ denotes the Poisson bracket between the functions $f,g$ on $M$. 

The one-parameter group of symplectomorphisms generated by $X_f$ will be denoted by $\psi_t^{f}$ whereas the induced one-parameter group of Poisson morphisms on $M$ will be indicated with  $\overline{\psi}_t^{f}$.  For $X_f$ a complete vector field,  the pair $(\psi_t^f, \overline{\psi}_t^{f})$ is for each $t$ a bundle isomorphism from $T^*M$ to itself.  As already mentioned, $\psi_t^{f}$ preserves the fibration of the vector bundle, but it does not act linearly among the fibers. Nevertheless, one can define a new linear structure on the fibers of the transformed vector bundle:
\begin{equation}
    \psi_t^{f}(p_x) +_f \psi_t^{f}(q_x) := \psi_t^{f}((p+q)_x)\,,\quad \lambda \cdot_f \psi_t^{f}(p_x) := \psi_t^{f}(\lambda \cdot p_x) 
\end{equation}
with $p_x, q_x\in T_x M$ and $\lambda\in \mathbb{R}$.

The one-parameter group of bundle isomorphisms $(\psi_t^f, \overline{\psi}_t^{f})$   induces transformations of the space of sections  of the vector bundle, $\Gamma(T^*M)$, which read \cite{Saunders}:
\begin{equation}\label{Ahat}
    \hat{A}^f_t (x) := (\hat{\psi}^{f}_t(A))(x) = \psi_t^{f}\circ A\circ \overline{\psi}_{-t}^{f}(x)\, \;\;\;\;\; A\in \Gamma(T^*M).
\end{equation}
The latter being a one-parameter group of transformations, it defines a vector field, $\tilde X_f(A)$, which is tangent to  $T^*M$ in $A(x)$ for each $x$,  $(\tilde X_f(A))_x\in T_{A(x)}T^*M$, namely, a vector field on the submanifold $\rm{im}(A)\subset T^*M$.

From \eqn{Ahat} we can read off the generator $\tilde X^f(A)$ to be \cite{Saunders} 
\be\label{Xtil}
(\tilde X_f(A))_x= X^f_{A(x)}- A_*(Y^f_x)
\ee
where only the first term survives when $X_f$ is vertical, namely in the standard case of vanishing $\Theta$. Notice that, in any case,  $\tilde X_f(A)$ is a vertical vector field, it being tangent to the fibre of the bundle.

On using Eqs. \eqn{localX}, \eqn{localY} we have 
\be\label{deltaA}
\tilde X_f(A) = (\gamma^\mu_\nu \frac{\del f}{\del x^\mu}-\Theta^{\mu\sigma}\frac{\del f}{\del x^\mu}\frac{\del A_\nu}{\del x^\sigma} )\frac{\del}{\del p_\nu}
\ee
with $\left\lbrace \frac{\del}{\del p_\nu}\right\rbrace$ a basis of vertical tangent  vectors in $TT^*M$. 
Therefore, the infinitesimal variation of the section $A$, associated with the symplectomorphism $X_f$,  is expressed as a vertical vector field along the image of the section $A$. This vector field has a geometrical interpretation: it is the generalized Lie derivative of the section $A$ with respect to the vector field $\tilde{X}_f$ which is projectable over $X_f$ (for more details see \cite{KMS_natural}, Chap.XI). 

From Eq. \eqn{deltaA} one recognizes the gauge transformation of the electromagnetic potential which has been found in \cite{Blumenhagen:2018kwq, Kupriyanov:2020sgx, Kupriyanov:2021cws} within different approaches and  it reproduces the standard gauge transformation of the potential for $\Theta\rightarrow 0, \gamma^\mu_\nu\rightarrow \delta^\mu_\nu$.  A conceptual difference of our derivation with respect to the  previous ones is that the infinitesimal gauge transformation of the potential emerges  here as  the component of a vertical vector field on the cotangent bundle.

Assuming the family of gauge transformations \eqn{Ahat}, and using the definition of $F$ \eqn{eq_1} as the pullback through the section $A$ of the symplectic form $\omega$, one  sees that the electromagnetic field $F(A)$ transforms as follows:
\begin{equation}\label{FAhat}
    F(\hat{A}^f_t) = (\psi_t^{f}\circ A\circ \overline{\psi}_{-t}^{f})^*(\omega) = (\overline{\psi}_{-t}^{f})^*\circ A^* \circ (\psi_t^{f})^* (\omega) = (\overline{\psi}_{-t}^{f})^*(F(A))\,, 
\end{equation}
where in the last step we have used the fact that $\psi_t^{f}$ is a one-parameter group of symplectomorphisms of the 2-form $\omega$. Once more, the previous formula reduces to the standard invariance of the electromagnetic field when the Poisson structure $\Theta$ vanishes. 

Eq. \eqn{FAhat} represents a one-parameter group of transformations of the vector bundle of $2$-forms over $M$, $\Omega^2 M$. Therefore, its infinitesimal generator  defines a vector field which is tangent to $\Omega^2 M$ in $F(x)$ for each $x$. It reads
\begin{equation}\label{Ftrans}
\begin{split}
    %\delta_f F 
    \tilde X_f(F) &= \left[[\left( \frac{\mathrm{d}}{\mathrm{dt}} (\overline{\psi}_{-t}^{f})^*(F(A)) \right)_{\mu \nu}\right]_{{t=0} }\frac{\partial}{\partial v_{\mu \nu}} = \left( \mathcal{L}_{Y_f}F(A) \right)_{\mu \nu} \frac{\partial}{\partial v_{\mu \nu}}  \\
    &=
    ( \left\lbrace f , F_{\mu \nu} \right\rbrace + F_{\sigma \nu} \partial_{\mu}(\Theta^{\rho \sigma}\partial_{\rho}f) - F_{\sigma \mu} \partial_{\nu}(\Theta^{\rho \sigma}\partial_{\rho}f) ) \frac{\partial}{\partial v_{\mu \nu}} 
\end{split}
\end{equation}
where $\mathcal{L}_{Y_f}F(A)$ is   the Lie derivative of the 2-form $F(A)$ along the vector field $Y_f=\Theta(\mathrm{d}f,\cdot)$. This is precisely  the infinitesimal variation under gauge transformation of the electromagnetic field, $\delta_f F $. Indeed,   the variation of the 2-form is described by a vertical vector field along the image of  $F(A)$, which is a section of $\Omega^2 (M)$. Accordingly, the vectors $\left\lbrace \frac{\partial}{\partial v_{\mu \nu}}\right\rbrace$ form a basis of vertical tangent vectors  to the vector bundle of 2-forms, $T\Omega^2 (M)$. 

We can conclude that  the variation of the electromagnetic field under a gauge transformation is defined by a derivation, the Lie derivative with respect to the vector field $\Theta(\mathrm{d}f,\cdot)$. Moreover,  the invariance of the electromagnetic field for $\Theta\rightarrow 0$ is recovered. 

 For comparison with previous derivations with similar results, in  \cite{Blumenhagen:2018kwq}  the  so-called $L_\infty$ bootstrap is adopted, which is  based on the  conjecture that Poisson (in fact fully noncommutative) gauge theories can be consistently constructed by completing an $L_\infty$ algebra, so that the action of gauge symmetries on fundamental fields is obtained by solving $L_\infty$ relations; then in \cite{Kupriyanov:2020sgx}  the same result is obtained by  requiring the closure of the algebra of gauge transformations with respect to the Poisson structure of spacetime. Finally, let us comment on the approach followed in   \cite{Kupriyanov:2021cws},  which is the one  that inspired our work. There, it is shown   that it is possible to derive the correct  gauge transformation for the potentials by means of a Poisson bracket on the cotangent space which is dictated by a symplectic embedding.   Elaborating on that,  we have clarified the geometric nature of the potentials  and of their gauge transformations at a global and local level,   as it emerges from   Eqs.  \eqn{Ahat}-\eqn{deltaA}. Then, the analysis is extended to  the   definition of the electromagnetic field and its gauge transformation \eqn{FAhat}, \eqn{Ftrans}.
As we will see, our  reformulation of the theory will naturally  bring to  the groupoid interpretation proposed in the second part of the paper. 

%which is a possible generalization of the standard gauge transformations of abelian and non-abelian Yang-Mills gauge theories 

\section{Lie Groupoids  and their  Lie Algebroids}\label{Lie Groupoids}
The theory of Lie groupoids and Lie algebroids is a natural generalization of the theory of Lie groups and Lie algerbas to the setting where the set of objects (see below) is not made of a single element, but forms a smooth manifold. In this section we are going to briefly review the main features of Lie groupoids and Lie algebroids which will be used in the second part of this paper. 

In particular, we are going to introduce the so called {\it action groupoid} associated with the action of a groupoid $G$ over a submersion map $\pi\,\colon\,S\,\rightarrow\,M$, where $M$ is also the space of objects of $G$. The corresponding Lie algebroid represents the infinitesimal version of this action over the same bundle. The relation with Poisson electrodynamics comes as follows: we will interpret the symplectic realization as a bundle over the base space $(M,\Theta)$ and the fields as sections of the latter. This bundle is endowed with the action of a suitable groupoid (the symplectic groupoid of the Poisson manifold $(M,\Theta)$), and the infinitesimal version of this action is expressed in terms of vector fields associated with differential 1-forms on $M$. Gauge transformations are expressed in terms of closed 1-forms, which generate via the exponential map suitable bisections acting on the sections of the symplectic realization.

Due to the wide literature on the subject, we will limit to a short summary of the mathematical structures involved, referring to \cite{Mackenzie} for the proofs of the main results and further details.

A groupoid $(G, M, s, t, 1_{\cdot}, \circ)$ is defined in terms  of the following structures:
\begin{itemize}
\item a pair of sets $G$ and $M$, the groupoid and the base, respectively, whose elements are  called morphisms (or arrows) for $G$   and objects (or units) for $\Omega$;
\item a pair of maps $s,t\,\colon\,G\,\rightarrow\,M$, with $s$ the source and $t$ the target;
\item a map $1_{\cdot}\,\colon\,M\,\rightarrow G$ called the object inclusion;
\item a partial multiplication $\circ\,\colon\, G^{(2)}\,\rightarrow \,G$ defined on the set 
$$
G^{(2)}=\left\lbrace (\beta,\alpha)\in G\times G\,\mid\, t(\alpha)=s(\beta) \right\rbrace
$$
of composable pairs.  
\end{itemize} 
Additionally, the composition law satisfies the following properties:
\begin{itemize}
\item associativity: for every triple of composable morphisms $(\gamma, \beta, \alpha)$ we have that $\gamma\circ(\beta\circ \alpha) = (\gamma \circ \beta)\circ \alpha$;
\item existence of units: for every morphism $\alpha\in G$ we have $1_{t(\alpha)}\circ \alpha = \alpha = \alpha\circ 1_{s(\alpha)}$;
\item existence of inverses: for every morphism $\alpha \in G$ there is an inverse morphisms $\alpha^{-1}\in G$ such that $\alpha^{-1}\circ \alpha = 1_{s(\alpha)}$ and $\alpha\circ \alpha^{-1}=1_{t(\alpha)}$. 
\end{itemize}
The notation $\alpha\,:\,x\,\rightarrow\, y$ will be occasionally employed to specify that the morphism $\alpha \in G$ has source $s(\alpha)=x$ and target $t(\alpha)=y$, whereas the groupoid $(G, M, s, t, 1_{\cdot}, \circ)$ will be denoted by $G \rightrightarrows M$, indicating only the groupoid and its base.  

Up to now we have introduced only the algebraic properties of the groupoid. A \textit{Lie groupoid} is a groupoid $G\rightrightarrows M$ where $G$ and $M$ are smooth manifolds such that the maps $s,t$ are surjective submersions, the object inclusion and the multiplication are smooth maps. The leaves $G_x = s^{-1}(x)$ and $G^x = t^{-1}(x)$ are submanifolds of the groupoid, whereas the morphisms $G_x^x = \left\lbrace \gamma \in G \,\mid\, s(\gamma)=t(\gamma)=x \right\rbrace$ form a Lie subgroup, called the isotropy subgroup at $x$.

A Lie group is a Lie groupoid, where the space of objects consists of a single point, the unit of the group. In this case every pair of morphisms is composable and has an inverse and, being $G$  a Lie group, the multiplication is a smooth map. 

The first non trivial example is the  {\it groupoid of pairs} of $M$,   $M\times M\rightrightarrows M$ whose morphisms are pairs $\alpha = (y,x)$ such that the source and the target maps are the projections onto the first and the second factor, respectively. The object inclusion is the diagonal inclusion and the multiplication is $(z,w)\circ (y,x) = (z,x)$ if and only if $y=w$, so that the inverse of the transition $\alpha$ is $\alpha^{-1}=(x,y)$. Therefore, the object inclusion and the multiplication are smooth maps whereas the source and the target are surjective submersions so that the axioms of a Lie groupoid are satisfied. 

Given a morhpism $\alpha\,\colon\,x\,\rightarrow\,y$ the right translation map is defined as the map $R_{\alpha}\,\colon\,G_y\,\rightarrow \,G_x$ such that $R_{\alpha}(\beta)=\beta\circ \alpha$, and an analogous definition holds for the left translation. Therefore, the right (resp. left) action of a groupoid on itself moves leaves of the source (target) map among themselves. Eventually, the inverse map $\tau\,\colon\,G\,\rightarrow\,G$, which associates with every morphism $\alpha$ its inverse $\alpha^{-1}$, is a diffeomorphism of $G$.

In order to introduce the infinitesimal description of a groupoid, we need to introduce a new geometrical structure, called Lie algebroid. A {\it Lie algebroid} $(A, \pi, M, a, [\cdot,\cdot])$ is defined in terms of  the following structures:
\begin{itemize}
\item a vector bundle $\pi\,\colon\,A\,\rightarrow\,M$ over a manifold $M$;
\item a vector bundle map $a\,\colon\,A\,\rightarrow\,TM$ called the anchor map;
\item a bracket $[\cdot,\cdot]\,\colon\,\Gamma(A)\times \Gamma(A)\,\rightarrow\,\Gamma(A)$ on the $C^{\infty}(M)$-module $\Gamma(A)$ of sections of the vector bundle $A$ which is $\mathbb{R}$-bilinear, alternating and it satisfies the Jacobi identity. 
\end{itemize}   
Additionally, the following compatibility properties must be satisfied:
\begin{itemize}
\item $[X,fY]= f[X,Y] + a(X)(f) Y$, where $a(X)(f)$ denotes the action by Lie derivative of the vector field $a(X)$ image of the section $X$ under the anchor map;
\item $a([X,Y]) = [a(X),a(Y)]$, which means that the anchor map is a Lie algebra homomorphism. 
\end{itemize}

A Lie algebra is a Lie algebroid where $M$ is a single point. Also the tangent bundle $\pi\,\colon\,TM\,\rightarrow\,M$ is a Lie algebroid with the usual bracket among vector fields and the anchor map is the identity map.

As in the case of Lie groups, the infinitesimal properties of a Lie groupoid $G\rightrightarrows M$ are described by a Lie algebroid $AG$ which is constructed out of the Lie groupoid. This construction requires several steps which are hereafter summarized. \\
 Firstly, one defines the vector bundle $\pi\,\colon\,AG \, \rightarrow\, M$ over the manifold $M$ whose fibres $\pi^{-1}(x)=:(AG)_x$ are the tangent spaces $T_{1_x}G_x$, where $T_{1_x}G_x$ denotes the tangent space to the leaf $G_x$ at the unit $1_x$. Secondly, a \textit{right-invariant vector field} on $G$ is defined as a vector field $X\in \mathfrak{X}(G)$ such that $s_*(X)=0$ and $(T_{\beta} R_{\alpha})(X(\beta)) = X(\beta\circ \alpha)$, where $T_{\beta} R_{\alpha}\,\colon\,T_{\beta}G_{t(\alpha)}\,\rightarrow \, T_{\beta\circ \alpha}G_{s(\alpha)}$ is the tangent map of the right translation. Analogously to the case of Lie groups, a right invariant vector field can be determined by its value on the submanifold of the units because we have that $X(\alpha) = (T_{1_{t(\alpha)}}R_{\alpha})(X(1_{t(\alpha)}))$. It can be proved that (see Sec.3.5 of \cite{Mackenzie}) there is an isomorphism between right invariant vector fields and sections of the vector bundle $AG$, which is actually an isomorphism of the associated modules $\mathfrak{X}^{RI}(G)$ and $\Gamma(AG)$ over the ring $C^{\infty}(M)$. Since $\mathfrak{X}^{RI}(G)$ is closed under the bracket of vector fields on $G$, a Lie bracket can be induced on the module $\Gamma(AG)$ of sections of the vector bundle $AG$. Eventually, the anchor map is induced from the map $t_*\,\colon\,TG\,\rightarrow\,TM$ restricted to the image of $AG$ as a subvector bundle of $TG$.

As examples of the previous construction we have the Lie algebra of a Lie group and the algebroid $TM$ associated with the groupoid of pairs $M\times M$.   
 
\subsection{Bisections of Lie groupoids and the exponential map}\label{Bisections}
In the theory of Lie groups Lie third theorem proves that, given a Lie algebra $A$, there is a unique simply connected Lie group $G$ for which $A$ is the associated Lie algebra. The same result is not true for general Lie algebroids, since there are obstructions to their integrability \cite{Crainic-fernandes_1}. 

Besides the concept of Lie groupoid integrating a Lie algebroid there is another construction which associates sections of the Lie algebroid $AG$ of a Lie groupoid $G$ to subsets of the Lie groupoid $G$, these are the {\it bisections}. As we will see, they  play an important role in generalizing the concept of finite gauge transformations within the groupoids approach to Poisson gauge theories. 

Let $G\rightrightarrows M$ be a groupoid over the base $M$. A \textit{bisection} $\sigma$ of $G$ is a map $\sigma\,\colon\,M\,\rightarrow \,G$ which is right-inverse to the source map $s$ and such that $t\circ\sigma \,\colon\,M\,\rightarrow\,M $ is a diffeomorphism. In other words the image of a bisection is a subset $B\subset G$ such that the source and target maps restricted to $B$ are diffeomorphisms. Given a bisection $\sigma$ one can construct another map $L_{\sigma}\,\colon\,G\,\rightarrow\,G$ defined as 
\begin{equation}
\label{eq:left-translation}
L_{\sigma}\alpha = \sigma(t(\alpha))\circ \alpha\,,
\end{equation}
which is a \textit{left-translation} of the groupoid $G$, i.e., a pair of maps $(\varphi,\varphi^0)$ with $\varphi\,\colon\,G\,\rightarrow\,G$ and $\varphi^0\,\colon\,M\,\rightarrow\,M$ such that $t\circ \varphi = \varphi^0\circ t$ and $s\circ \varphi = s$, and such that each $\varphi^x\,\colon\,G^x \,\rightarrow\, G^{\varphi^0(x)}$ is a left translation by a unique element of the groupoid $\alpha\,\colon\,x\,\rightarrow\,\varphi^0(x)$. 

The set $\mathcal{B}$ of all bisections of a groupoid is a group under the multiplication
\begin{equation}
(\rho \star \sigma)(x) = \rho(t(\sigma(x)))\circ \sigma(x)
\end{equation}  
with identity being the object inclusion and inverses given by $\sigma^{-1}(x) = \sigma((t\circ \sigma)^{-1}(x))^{-1}$.

 One can also consider \textit{local bisections}, i.e., maps $\sigma_U\,\colon\,U\,\rightarrow\,G$ which are right inverse to the source map $s$ and such that $t\circ \sigma_u$ is a diffeomorphism from $U$ to the subset $t(\sigma_U(U))\subset \Omega$. Analogously there are local left-translations and the map $\eqref{eq:left-translation}$ associates a local left-translation to every local bisection. More generally the following theorem holds:
\begin{theorem}\label{thm1} {\rm (\cite{Mackenzie} sec 1.4)}
Let $G\rightrightarrows M$ be a Lie groupoid,  with $\varphi\,\colon\,\mathcal{U}\,\rightarrow\,\mathcal{V}$  a diffeomorphism from $\mathcal{U}\subseteq G$ to $\mathcal{V}\subseteq G$.  Let $f\,\colon\,U=t(\mathcal{U})\,\rightarrow \,V=t(\mathcal{V})$ be a diffeomorphism from $U$ to $V$ such that $s\circ \varphi = s$ and $t \circ \varphi = f \circ t$ and $\varphi(\beta\circ \alpha) = \varphi(\beta)\circ \alpha$ whenever $(\beta,\alpha)\in G^{(2)}$ and both $\beta$ and $\beta\circ \alpha$ belong to $\mathcal{U}$. Then $\varphi$ is the restriction to $\mathcal{U}$ of a local left-translation $L_{\sigma_U}\,\colon\,G^{U}=t^{-1}(U)\,\rightarrow\,G^{V}=t^{-1}(V)$ associated to a local bisection $\sigma_U$. The set of all local bisections on $U$ will be denoted by $\mathcal{B}_U(G)$. 
\end{theorem}  

Let us now come back to the Lie algebroid $\pi\colon AG\,\rightarrow M$ of a Lie groupoid $G\rightrightarrows M$ introduced in the previous subsection. A section of $\mathcal{X} \in \Gamma(AG)$ determines a right invariant vector field ${X}\in \mathfrak{X}^{RI}(G)$ so that there is a local flow $\varphi_t \,\colon\,\mathcal{U}\,\rightarrow\,\mathcal{U}_t$ on $G$.   Since ${X}$ is right invariant, we have that $s\circ \varphi_t = s$ (because it is tangent to the leaves of the source map) and $R_{\beta} \circ \varphi_t  = \varphi_t\circ R_{\beta}$ for every $\beta \,\colon\,x\,\rightarrow\,y$ such that $\mathcal{U}\cap G_x\neq \emptyset$ and $\mathcal{U}\cap G_y \neq \emptyset$ and for all $t$. Thanks to the  previous properties, one can prove that there is an associated local flow $\psi_t\,\colon\,t(\mathcal{U})\,\rightarrow\,t(\mathcal{U}_t)$ which is generated by the vector field $t_*({X})=a(X)$, where $a$ is the anchor map of $AG$. 

Because of the above discussion, we have that the local flows $(\varphi_t,\psi_t)$ satisfy the properties of Theorem \ref{thm1}, so that each $\varphi_t$ is the restriction to $\mathcal{U}$ of a local left translation $L_{\sigma_t}$ associated with the local bisections $\sigma_t$. Therefore, the following theorem holds:
\begin{theorem}\label{thm2} {\rm(\cite{Mackenzie} sec 1.4)}
Let $G\rightrightarrows M$ be a Lie groupoid, $W\subseteq M$ an open subset of $M$, and  $\mathcal{X}\in \Gamma_W(AG)$ a local section of the Lie algebroid $AG$. Then, for each $x_0\in M$ there is an open neighborhood $U$ of $x_0$ in $W$, an $\epsilon >0$ and a unique smooth family of local bisections $\mathrm{Exp} \,t\mathcal{X} \in \mathcal{B}_U(G)$, $-\epsilon < t < \epsilon$, such that 
\begin{itemize}
\item $\frac{d}{dt}\mathrm{Exp}\,t\mathcal{X} \,\mid_{t=0} = \mathcal{X}$;
\item $\mathrm{Exp}\,0\mathcal{X} = 1_U $, where $1_U$ is the identity local bisection on $\mathcal{U}$;
\item $\mathrm{Exp}(t+s)\mathcal{X} = \mathrm{Exp}\,t\mathcal{X}\star \mathrm{Exp}\,s\mathcal{X}$;
\item $\mathrm{Exp}(-t\mathcal{X}) = (\mathrm{Exp}\,t\mathcal{X})^{-1}$;
\item $t(\mathrm{Exp}(t\mathcal{X}))$ is a local 1-parameter group of transformations for $a(\mathcal{X})\in \mathfrak{X}_U(\Omega)$, where $\mathfrak{X}_U(\Omega)$ is the space of local sections of $T\Omega$ on $U\subseteq \Omega$. 
\end{itemize}  
\end{theorem} 
The map $\Phi_{\mathcal{X}}\,\colon\,\mathbb{R}\times U \,\rightarrow \,G$ defined as $\Phi_\mathcal{X}(t,x)=(\mathrm{Exp}\,t\mathcal{X})(x)$ is smooth as a consequence of the smoothness of the flow of a smooth vector field. Therefore there is an \textit{exponential map} which associates with any local section $\mathcal{X}$ of the algebroid $AG$ of a groupoid $G$ a local bisection $\sigma = \mathrm{Exp}\,\mathcal{X} = \Phi(1,\cdot)$. In this way, one can interpret local bisections of $G$ as generalized elements of the groupoid. 

From now on we shall use the same symbol $X$ to denote the sections of $AG$ and the associated vector fields on $G$. 

\subsection{Action groupoids and their  algebroids}\label{Action groupoids}
In this subsection we will present a family of groupoids which naturally arise when dealing with field theories over space-time manifolds endowed with suitable additional structures. These are the action groupoids. We will show in particular  that they provide a natural generalization of the group of gauge transformations for Poisson gauge theories. 

Let us consider  a smooth submersion of the spacetime manifold $M$,  $\pi\,\colon\, S \,\rightarrow \, M$. Let $G\rightrightarrows M$ be a Lie groupoid over $M$ whose transitions represent the symmetries of the spacetime under consideration. For instance in the case of  $M$ being endowed with a metric tensor $g$, one can consider the so called Riemannian frame groupoid (\cite{Mackenzie} sec 1.7) whose transitions are linear invertible transformations $T_{xy}$ from $T_xM$ to $T_yM$ such that $T_{xy}^*g_y = g_x$. 

On introducing  $G\ast S$, the pullback manifold 
$$
G\ast S = \left\lbrace (\alpha, \xi) \in G\times S \mid s(\alpha) = \pi(\xi)\right\rbrace\,.
$$
we consider an action of the groupoid $G\rightrightarrows M$ on the bundle $S\rightarrow M$, i.e., a smooth map $\Phi\,\colon\,G\ast S \,\rightarrow$ such that:
\begin{itemize}
\item $\pi(\Phi(\alpha,\xi)) = t(\alpha)$;
\item $\Phi(\beta, \Phi(\alpha,\xi)) = \Phi(\beta\circ \alpha, \xi)$ for every $(\beta,\alpha) \in G^{(2)}$
\item $\Phi(1_{\pi(\xi)},\xi) = \xi$ for every $\xi \in E$.
\end{itemize}

Given $(G\ast S, \Phi)$ one can build the associated action groupoid as follows:
\begin{itemize}
\item the set of morphisms is $G\ast S$ while the base is $S$;
\item the source is the projection onto the factor $S$, while the target map is the action $\Phi$;
\item the object inclusion is given by the map which associates the transition $\Phi(1_{\pi(\xi)},\xi)$ with the point $\xi \in E$;
\item given two transitions $(\beta, \eta)$ and $(\alpha,\xi)$ with $\eta = \Phi(\alpha, \xi)$, the composition is defined as $(\beta,\eta)\circ (\alpha,\xi) = (\beta\circ \alpha, \xi)$;
\item given the transition $(\alpha,\xi)$ its inverse is the transition $(\alpha^{-1}, \Phi(\alpha,\xi))$.
\end{itemize}
With the above structure $G\ast S\,\rightrightarrows \,S$ is a Lie groupoid over the base $S$.

Analogously, let $\pi_A\,\colon\,A\,\rightarrow \,M$ be a Lie algebroid with the corresponding bracket denoted by $[,]_A$ and $\pi\,\colon\,S\,\rightarrow\,M$ a surjective submersion defining a bundle over $M$. An action of the Lie algebroid $A$ over the map $\pi$ is a map $\varphi\,\colon\,\Gamma(A)\,\rightarrow\,\mathfrak{X}(S)$ such that:
\begin{itemize}
\item $\varphi(X+Y)=\varphi(X)+\varphi(Y)$;
\item $\varphi(f X) = (\pi^*(f))\varphi(X)$, where $f\in C^{\infty}(M)$;
\item $\varphi([X,Y]_A)=[\varphi(X),\varphi(Y)]$, where the bracket on the r.h.s. is the Lie bracket of vector fields;
\item $\varphi(X)$ is $\pi$-related to $a(X)$, where $a\,\colon\,A\,\rightarrow\,TM$ is the anchor of the Lie algebroid. 
\end{itemize}  
Given the pullback bundle $\pi^*(A)$ which is a vector bundle over the base $S$, one can define a Lie algebroid structure as follows:
\begin{itemize}
\item the anchor $a'\,\colon\,\pi^*(A)\,\rightarrow\,TM$ acts on a section $Y = \sum_{j} f_j X_j\in \Gamma(\pi^*(A))$, with $f_j \in C^{\infty}(S)$, as $a'(Y)=\sum_j f_j \varphi(X_j)$;
\item $[Y_1, Y_2]_{A'} = [\sum_{j} f_j X^{(1)}_j, \sum_{k} g_k X^{(2)}_k]_{A'} = \sum_{j,k}f_jg_k[X^{(1)}_j, X^{(2)}_k]_{A}+\sum_{j,k} f_j\varphi(X^{(1)}_j)(g_k) X^{(2)}_k - \sum_{j,k} g_k\varphi(X^{(2)}_k)(f_j) X^{(2)}_j$\,.
\end{itemize}
The Lie algebroid obtained in this way will be denoted by $\pi^!A$ and it is a Lie algebroid over the base $S$.

These two constructions are related. Indeed, given the action groupoid $G\ast S$ with the associated action $\Phi$, the corresponding Lie algebroid is isomorphic to the Lie algebroid $AG\,\rightarrow\,M$ of the Lie groupoid $G\rightrightarrows M$ with an action over the map $\pi\,\colon\,S\,\rightarrow \,M$ given by 
$$
\varphi(X)(\xi) = T_{1_{\pi(\xi)}}\Phi_{\xi}(X(\pi(\xi)))\,.
$$
In the formula above, $X\in \Gamma(AG)$ is a section of $AG$ whereas $\Phi_{\xi}\,\colon\,G_{\pi(\xi)}\,\rightarrow\,S$ is the evaluation of the map $\Phi$ at the point $\xi\in S$, i.e., $\Phi_{\xi}(\alpha)= \Phi(\alpha, \xi)$.

Finally, let us consider $B\subset G$, the image of a bisection   of the Lie groupoid $G\rightrightarrows M$,   (its source and target map will be denoted by $s_G,t_G$), i.e., the graph of a map $\sigma_B\,\colon\,M\,\rightarrow\,G$ such that $\overline{\Phi}_B = t_G\circ \sigma_B \,\colon\,M\,\rightarrow\,M$ is a diffeomorphism. Then, one can define a bisection of the action groupoid $G\ast S \rightrightarrows S$ as the subset $B_S =\left\lbrace (\alpha, \xi)\in B\times S \,\mid\, \pi(\xi)=s(\alpha) \right\rbrace$ which is the graph of the pullback $\pi^*\sigma_B$ of the bisection $\sigma_B$, i.e., $\pi^*\sigma_B\,\colon\,S\,\rightarrow\,G\ast S$ given by 
$$
\pi^*\sigma_B(\xi)=(\sigma_B(\pi(\xi)), \xi)\,. 
$$
The diffeomorphisms $\Phi_B\,\colon\,S\,\rightarrow\, S$ induced by these bisections, i.e., the maps $\Phi_B = t\circ \pi^*\sigma_B$ given by 
$$
\Phi_B(\xi) = \Phi(\sigma_B(\pi(\xi)),\xi)\,,
$$
map the fibre $\pi^{-1}(\pi(\xi))$ to the fibre $\pi^{-1}(\overline{\Phi}_B(\pi(\xi)))$. Therefore, they determine a bundle morphism, which, in general, does not need to preserve additional structures along the fibres. We will see in the next sections how these structures are related with   the gauge transformations of Sec. \ref{Gauge transformations 1}.

\section{Back to Poisson Electrodynamics} \label{back to ED}
In this section we come back to Poisson electrodynamics and we interpret the description outlined in Sec. \ref{Poisson Electrodynamics} in the light of the geometrical properties of Lie groupoids and algebroids. Indeed, groupoids and algebroids generalize, respectively, the global and local properties of group and algebra actions using the abstract language of categories. Adopting groupoids and algebroids as models for the description of field theories, one can use bisections of groupoids and sections of algebroids as kinematical or structural ``gauge'' symmetries of the model. We think that this clarifies  some of the features of Poisson electrodynamics which  differentiate them from standard Yang-Mills gauge theories. As already mentioned in the introduction, the connection between groupoids and gauge transformations in Poisson electrodynamics was already pointed out in \cite{Kupriyanov:2023qot}. Many results obtained in this section could be read as local approximations of the results presented in \cite{Kupriyanov:2023qot}, where the focus is on the description of finite gauge transformations. However, we want to stress once more that from our point of view, bisections of groupoids play a role only from the point of view of gauge transformations. Fields, on the other hand, are sections of a local symplectic realization of the Poisson spacetime $M$. In our approach, the fact that this manifold can be endowed with the structure of a groupoid does not play a role as far as the fields are concerned. 

\subsection{Potentials in the Lie Algebroid setting}
In this framework the space of fields, namely  the  potential one forms, is the set of   sections of the vector bundle $T^*M$. For $M$ a Poisson manifold, $T^*M$ can be endowed with a Lie algebroid structure where the anchor $a\,\colon\,\Gamma T^*M\,\rightarrow \, \mathfrak{X}(M)$ and the Lie bracket $\left[ \cdot,\,\cdot \right]_{\Theta}\,\colon\,\Gamma T^*M\times \Gamma T^*M \,\rightarrow\, \Gamma T^*M$ are defined as follows:  
\begin{equation}\label{cotgalgebroid}
\begin{split}
    a(A) &= \Theta(A,\,\cdot ) \\
    [A_1,\,A_2]_{\Theta} &= \mathcal{L}_{a(A_1)}A_2 - \mathcal{L}_{a(A_2)}A_1 - \mathrm{d}(\Theta(A_1,\,A_2))\,.
\end{split}
\end{equation}
Ignoring for the scopes of this paper  situations where topological obstructions can occur, the  Lie algebroid  $(T^*M,\,a,\,\left[\cdot,\,\cdot \right]_{\Theta})$ can be integrated to  a Lie groupoid, the so-called symplectic groupoid,  $\Sigma(M)$ (the topological obstructions refer to the possibility of endow a certain groupoid with a smooth structure). $\Sigma(M)$ is the homotopy groupoid associated with the foliation generated by the Hamiltonian vector fields of the Poisson manifold $(M,\Theta)$ and carries a canonical symplectic structure $\omega$. Namely,  its morphisms  are appropriate homotopy equivalence classes of paths which live  on the symplectic leaves generated by the Hamiltonian vector fields on $M$  (for more details see \cite{Crainic-fernandes_2}). A different construction of the symplectic groupoid is obtained as the space of solutions of the Poisson sigma-model\cite{cattaneo-felder_2001}.

Other than that, the manifold $(\mathcal{U}\subset T^*M,\omega)$ is a symplectic realization of the Poisson manifold $(M;\Lambda)$ with dual orthogonal symplectic foliations. As described in \cite{Crainic-fernandes_2} complete symplectic realizations of a Poisson manifold exist if and only if the Poisson manifold is integrable, in the sense that the Lie algebroid $(T^*M,\,a,\,\left[\cdot,\,\cdot \right]_{\Theta})$ integrates to the associated symplectic Lie groupoid. In general, sections of the bundle $\pi\,\colon\,\mathcal{U}\,\rightarrow\,M$ are 1-forms on $M$ which are only defined  in the neighborhood $\mathcal{U}$ of the zero section.  %This is related to the fact that 

A symplectic realization of $(M, \Theta)$ be it complete or not,  %$\pi\,\colon\,(S,\omega)\,\rightarrow\,(M,\Theta)$  
carries a representation of the Lie algebroid $(T^*M,\,a,\,\left[\cdot,\,\cdot \right]_{\Theta})$ that can be integrated. The action of the Lie algebroid on the sections of the symplectic realization is  given by the map
\begin{equation}
    \varphi(\mathrm{d}f)= X_{\pi^*(f)}\,,
\end{equation}
where $\mathrm{d}f, f\in C^{\infty}(M)$ are  sections of the Lie algebroid $T^*M$ which are closed, hence locally exact  one forms, and $X_{\pi^*(f)}\,\in \, \mathfrak{X}(\mathcal{U})$ are the Hamiltonian vector fields on $\mathcal{U}$ associated with the pull-back $\pi^*(f)\in C^{\infty}(\mathcal{U})$  (see \cite{Crainic-fernandes_2} for further details). 

Therefore, gauge potentials  of Poisson electrodynamics can be identified with the sections of the local symplectic realization $\mathcal{U}\in T^*M$, which, moreover,  carries an action of the Lie algebroid $T^*M$ associated with the Poisson manifold $(M,\Theta)$. 

\subsection{Gauge transformations}
Let us see how the action of the algebroid  $T^*M$  is precisely the one which  encodes  infinitesimal  gauge transformations.

We have seen that the symplectic realization $\mathcal{U}$ carries an action of the Lie algebroid $T^*M$ which induces an action on the space of sections of the symplectic realization. In particular,  given a section $\chi\in \Omega^1(M)$ of the Lie algebroid $T^*M$, which in addition is a closed (and locally exact) 1-form, one can define the vector field $\varphi(\chi)\in \mathfrak{X}(\mathcal{U})$ which is exactly the vector field $X_{f}$ defined in Eq.\eqref{Xf}, since locally $\chi = \mathrm{d}f$ for some $f\in C^{\infty}(M)$. Following now the same steps as in Sec. \ref{Gauge transformations 1} one arrives at the action on the space of sections of $\mathcal{U}\subset T^*M$ given by Eq.\eqref{Ahat}. 

To be more precise, the exponential map of the algebroid $T^*M$ associates with the section $\mathrm{d}f \in \Omega^1(M)$ a 1-parameter group of bisections of the symplectic groupoid $\Sigma(M)$, say $\mathrm{Exp}(t\,\mathrm{d}f)$. Then, one can construct the pullback of this 1-parameter group  to the action groupoid $\Sigma(M)\ast \mathcal{U}$ defined in   Sec. \ref{Action groupoids},  obtaining a new 1-parameter group of bisections, $\pi^*(\mathrm{Exp}(t\,\mathrm{d}f)$. Then, the composition of the latter with the target map of the action groupoid gives a 1-parameter group of diffeomorphisms of $\mathcal{U}$, which is precisely the one  defined in Sec.\ref{Gauge transformations 1}, $\psi_t^f\,\colon\,\mathcal{U}\,\rightarrow\,\mathcal{U}$. As already explained, $\psi_t^f$ preserves the bundle structure of the symplectic realization $\mathcal{U}$ so that  we can extend its action to the space of sections of   $\mathcal{U}$. The generator of the 1-parameter group of transformations has  been introduced in Eq.\eqref{deltaA} and it has been interpreted as the generator of the gauge transformations of  Poisson electrodynamics. Let us stress once more that in the limit of vanishing $\Theta$ one recovers the group of gauge transformations of the standard electromagnetism. When $\Theta\neq 0$ the gauge transformations involve both space-time and momenta and momenta are transformed in a nonlinear way. 

The previous description can be thought as a first order approximation of a complete theory, where by  complete we mean that fields are sections of a complete symplectic realization $\pi\,\colon\,(S,\omega)\,\rightarrow\,(M,\Theta)$ of the Poisson manifold $(M,\Theta)$. If one requires that the action of the symplectic groupoid $\Sigma(M)$ on $S$ determine a principal $\Sigma(M)$-bundle over the quotient $L = S/\Sigma(M)$ (which however is not  a manifold in general), the orbit space will carry a canonical Poisson structure $\Theta_L$ whose symplectic leaves are the symplectic manifolds $\pi^{-1}(x)/\Sigma(M,x)$, where $\Sigma(M,x)$ is the isotropy group of the symplectic groupoid at the point $x\in M$. Since $S$ is a symplectic realization also for $(L,\Theta_L)$, one can consider the symplectic groupoid $\Sigma(L)$ associated with $(L,\Theta_L)$ and repeat the same procedure as before. The quotient $S/\Sigma(L)$ will give back the base Poisson manifold $(M,\Theta)$. This description suggests that Poisson gauge theories come in dual pairs, and if one chooses the symplectic realization given by the symplectic groupoid $\Sigma(M)$ with its symplectic form, the dual Poisson manifold $(L,\Theta_L)$ is anti-isomorphic to the manifold $(M,\Theta)$. This  interesting duality of Poisson gauge theories seems to  be related with Born reciprocity and $T$-duality, as already noticed in \cite{Kupriyanov:2023qot}. We plan to come back to the subject within a complete formulation of the theory.

Before closing the section, let us come back to the action of the gauge transformations on the sections of the local symplectic realization $\mathcal{U}\subset T^*M$. As we have already noticed, this action is not linear on the fibers. However, one can see that it provides  transformations among non linearly related realizations of the same abstract Lie algebroid. Indeed, as already illustrated, using fiber-preserving diffeomorphisms it is possible to define new vector bundle structures on $T^*M$ depending on the gauge choice. Analogously, one could define a transformed bracket $\left[\cdot,\,\cdot  \right]_f$ and anchor $a_f$ as follows;
\begin{equation}
\begin{split}
    \left[ \hat{A}^f_t\,,\hat{B}^f_t\right]_f = \widehat{\left[ A\,,B\right]}^f_t\,,\\
    a(\hat{A}^f_t) = (\overline{\psi}^f_t)_*\Theta(A\,,\cdot)\,.
\end{split}
\end{equation}  
Interestingly, a similar class of equivalence of Lie algebroids appears when dealing with corner symmetries in gauge theories \cite{KLP23}. The precise relation among the two approach is  another interesting issue  that we plan to investigate in a future publication. 

\subsection{Covariance and Invariance in the groupoid interpretation of Poisson electrodynamics}
The sections of the symplectic realization $T^*M$ have been used  to pull-back the symplectic two-form $\omega$ and get the electromagnetic field $F(A)$. As already mentioned, this  can be seen as a local version of a complete description  of Poisson electrodynamics in terms of complete symplectic realizations. 
The definition of $F$ as the pull-back $A^*(\omega)$
furnishes an electromagnetic  two-form which transforms covariantly under gauge transformations, as has been shown  in Sec. \ref{Gauge transformations 1}. On the other hand, we have seen that the symplectic realization $\mathcal{U}$, as well as every complete symplectic realization of the Poisson manifold $(M,\Theta)$, is endowed with a pair of foliations, the first one is  $\mathcal{F}(\pi)$,  whose vector fields are tangent to the leaves of the bundle map $\pi\,\colon\,\mathcal{U}\,\rightarrow\,M$, while  the other one,  $\mathcal{F}(\pi_1)$,  is the symplectic orthogonal of the previous foliation. The projection $\pi_1:\mathcal{U}\,\rightarrow M$ provides the symplectic realization of the dual Poisson manifold, which is anti-isomorphic to the Poisson manifold $(M,\Theta)$. Therefore, a section $A^{(1)}$ of the bundle $\pi_1\,\colon\,\mathcal{U}\,\rightarrow\,M$ can be used to pullback the symplectic two-form $\omega$ to obtain a different closed two-form on $M$, say $F^{(1)}(A)$. Since the generators of gauge transformations are vectors which are tangent to the leaves of the foliation $\mathcal{F}(\pi_1)$, the gauge transformations are transformations which preserve the leaves of the bundle $\pi_1\,\colon\,\mathcal{U}\rightarrow\,M$ and reduce to the identity on the base manifold $M$. Therefore, repeating the computations in Sec.\ref{Gauge transformations 1}, one can see that the two-form $F^{(1)}(A)$ is {\it invariant} under these transformations. However, this should be interpreted as a dual electromagnetic field. As already mentioned in the introduction, this feature of Poisson electrodynamics was already described in ref.\cite{Kupriyanov:2023qot}, where the two electromagnetic fields are defined using the two maps that are associated with any bisection of the symplectic groupoid $\Sigma(M)$. Forthcoming investigations are required to properly understand the origin and the meaning of this duality.

%\section{``Minimal Coupling'' revised}

\section{Conclusions and Perspectives}\label{conclusions}
In this paper we have proposed an interpretation of Poisson electrodynamics in terms of Lie algebroids and Lie groupoids actions. The description obtained here  is a first-order approximation of a complete theory, whose fields will be generalized elements of a suitable Lie groupoid. This approach fits well in the groupoidal approach to field theories which some of the authors have been developing \cite{CDCIMSZ-2021a,CDCIMSZ-2021b}. According to it, fields are generalized histories on a suitable Lie groupoid and the associated von Neumann groupoid algebra will determine the observables of the theory. Therefore, the first-order approximation of these fields are (space-time dependent) sections of the corresponding Lie algebroid. The exponential map allows to recover the fields from their first-order approximations: This correspondence allows also to induce a $\star$-product on the space of functions on the Lie algebroid. 

The results contained in this paper are preliminary and further investigation is required. In particular, only fields and ``kinematical'' gauge transformations have been defined. In order to introduce a suitable dynamics, one can proceed in analogy with  standard electrodynamics. If the space-time is endowed with a metric structure which is invariant under gauge transformations,  a  Lagrangian  can be defined as 
\begin{equation}
\mathcal{L}(F)=\int_{M}F(A)\wedge \star F(A)\,,
\end{equation}    
where the symbol $\star$ here denotes  the Hodge-star operation on forms. Then, a generalized Ampère law will be obtained but, in general, only part of the kinematical gauge group will be also a ``dynamical'' group of gauge transformations. Dynamical Lagrangian functions have been proposed in \cite{Kupriyanov:2023gjj} for specific linear Poisson structures with a slightly different   definition of the electromagnetic field with respect to the one proposed in this paper. For those Lagrangians, an analysis of the constraints in the Hamiltonian formalism as been performed in \cite{Bascone:2024mxs} in order to show that the constraints are first class, hence generating gauge transformations. Other possible Lagrangian functions have been also discussed in ref.\cite{Kupriyanov:2023qot}. A careful analysis has to be performed in order to compare all these models. 

A second aspect which deserves further investigation is the definition of a coupling with matter fields. From the analysis presented in the last section one notices that the action of an algebroid is present on the symplectic realization, which integrates to a representation of the associated Lie groupoid. Therefore, one could consider  matter fields as sections of a vector bundle carrying a linear representation of the symplectic groupoid $\Sigma(M)$: the action of the corresponding Lie algebroid will provide a suitable generalization of the minimal coupling for the standard electrodynamics. This action will be represented via a suitable derivation of the vector bundle, which requires an algebroid connection (see \cite{Crainic-fernandes_1} for further details). A Lagrangian describing the coupling to a particle was proposed in ref.\cite{Kupriyanov:2023qot}, using bisections (gauge potentials) to modify the Hamiltonian function of the particle in absence of the field. One could also extend this approach to more general matter fields, but we think this should be thought within the multisymplectic formalism for the description of field theories (see for instance \cite{CDCIMSZ-2022} for some recent applications of the formalism).

Another important aspect which needs to be addressed is the extension of this approach to non-Abelian gauge theories.  %there are different ways of extending the analysis outlined in this paper to the case when non-Abelian charges are present. 
One possibility would be to consider $\mathfrak{g}$-valued one-forms on $T^*M$, where $\mathfrak{g}$ is the Lie algebra of a Lie group defining the charges of the field. If a  one-form  of this kind defines a connection, one could consider its pullback along a section $A$ of $T^*M$ to define a connection 1-form on the spacetime $M$. As explained previously, the gauge transformations of Poisson electrodynamics can be interpreted as a representation of the Lie algebroid $T^*M$ on the space of fields. Therefore, one could define analogous gauge transformations if the space of $\mathfrak{g}$-valued 1-forms would possess a Lie algebroid structure. 
We shall come back to this and previously mentioned  issues in a forthcoming work.

\bibliographystyle{utphys}
\bibliography{ref}

\end{document}